# Towards Three-Dimensional Weyl-Surface Semimetals in Graphene Networks

Chengyong Zhong[1], Yuanping Chen[1*], Yuee Xie[1*], Shengyuan A. Yang[2], Marvin L. Cohen[3] and Shengbai Zhang[4*]

[1]School of Physics and Optoelectronics, Xiangtan University, Xiangtan, Hunan 411105, China

[2]Research Laboratory for Quantum Materials, Singapore University of Technology and Design, Singapore 487372, Singapore

[3]Department of Physics, University of California at Berkeley, and Materials Sciences Division, Lawrence Berkeley National Laboratory, Berkeley, California 94720, USA

[4]Department of Physics, Applied Physics, and Astronomy Rensselaer Polytechnic Institute, Troy, New York 12180, USA

**Abstract:** Graphene as a two-dimensional (2D) topological Dirac semimetal has attracted much attention for its outstanding properties and potential applications. However, three-dimensional (3D) topological semimetals for carbon materials are still rare. Searching for such materials with salient physics has become a new direction for carbon research. Here, using first-principles calculations and tight-binding modeling, we study three types of 3D graphene networks whose properties inherit those of Dirac electrons in graphene. In the band structures of these materials, two flat Weyl surfaces appear in the Brillouin zone (BZ), which straddle the Fermi level and are robust against external strain. When the networks are cut, the resulting lower-dimensional slabs and nanowires remain to be semimetallic with Weyl line-like and point-like Fermi surfaces, respectively. Between the Weyl lines, flat surface bands emerge with strong magnetism when each surface carbon atom is passivated by one hydrogen atom. The robustness of these structures can be traced back to a bulk topological invariant, ensured by the sublattice symmetry, and to the one-dimensional (1D) Weyl semimetal behavior of the zigzag carbon chain, which has been the common backbone to all these structures. The flat Weyl-surface semimetals may enable applications in correlated electronics, as well as in energy storage, molecular sieve, and catalysis because of their good stability, porous geometry, and large superficial area.

# Introduction

Carbon is one of the most abundant elements in the universe. Besides graphite and diamond, carbon has many allotropes because its atoms can be bonded together strongly in different states of hybridization (sp, $sp^2$ and $sp^3$)[1,2]. In recent decades, important phenomena discovered in various carbon allotropes (from fullerene to carbon nanotube to graphene) have resulted in a great deal of exciting researches on these materials, for their fascinating properties[3-9] and potential applications[10,11].

One of the properties of graphene is its topological semimetal character, i.e., the Fermi surface consists of two Fermi points where the conduction and valence bands touch with linear energy dispersion[12,13]. It is noted that although commonly known as a Dirac semimetal, graphene is in fact also a two-dimensional (2D) "Weyl semimetal" because the negligibly small spin-orbit coupling (SOC) strength makes it possible to treat the electron spin as a dummy variable. The unique band structure of graphene gives rise to a giant electron mobility and other exotic physical phenomena such as the abnormal quantum Hall effect[14]. However, graphene is only a single-atomic-layer sheet, which is suitable for planar electronics but difficult to be integrated with other higher-dimension devices[15-17]. Multi-layer nanosheets obtained by van der Waals epitaxy cannot resolve this issue, because the weak interaction between layers may hamper their electronic communication. Therefore, it is desirable to have three dimensional (3D) carbon allotropes to extend the Dirac/Weyl physics to higher dimensions. In this context, most of the 3D carbon allotropes proposed to date, such as $sp^3$-hybridized M-carbon[18], CKL[19], $sp^2$-hybridized H6-carbon[20], bct-4[21], and $sp^2/sp^3$-hybridized T6-carbon[22], Squaroglitter[23], are not semimetals.

Recent experimental and theoretical studies indicated that graphene networks consisting of interconnected graphene or graphene nanoribbons maybe good candidates of 3D Dirac/Weyl semimetals. For example, an interconnected graphene networks grown by chemical vapor deposition show a very high electrical conductivity which could be six orders of magnitude higher than that of the chemically-derived graphene-based composites[24]. A graphene network, termed Mackay-Terrones crystal, is a topological nodal-line semimetal, based on first-principle calculations[25]. An interpenetrated graphene network we studied recently also possesses topological nodal lines at the Fermi surface, which will evolve into a Weyl semimetal after breaking the inversion symmetry[26]. More general, graphene networks such as carbon foams are formed by linking graphene nanoribbons[27-29].

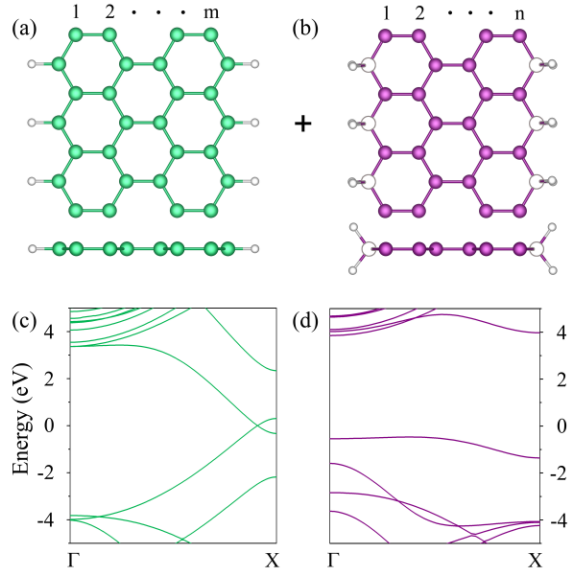

Figure 1. (a-b) Two elemental building blocks of 3D graphene networks, which are both zigzag graphene nanoribbons. The widths of (a) the green and (b) purple colored graphene nanoribbons are their respective numbers of zigzag chains, labeled here as m and n. Upon linking, atoms at the two edges of the purple colored nanoribbon (hollow circles) are $sp^3$-hybrdizated. (c-d) Band structures of the two zigzag graphene nanoribbons in (a) and (b) with m = 2 and n = 2, respectively.

Here we consider the linkage of zigzag nanoribbons, where two basic kinds of ribbons with different physical properties emerge: in the first kind, all (green) carbon atoms are $sp^2$, as shown in Fig. 1(a). In the second kind, however, while all the (purple) carbon atoms remain to be $sp^2$, all the edge (white) carbon atoms become $sp^3$, as shown in Fig. 1(b). Three families of graphene networks (shown in Fig. 2) can be constructed this way, where two green nanoribbons are always connected together by a purple nanoribbon. Figures 1(c-d) show that the green nanoribbon is a one-dimensional Weyl semimetal, while the purple nanoribbon is a semiconductor. These drastically different electronic properties are due solely to their different edge passivation chemistry. In such graphene networks, should topological properties exist, they are the results of green nanoribbons. However, the purple nanoribbons not only provide the backbones for the stability of the overall structure, but they also determine the semiconducting behavior when carrier transport takes place in directions orthogonal to that of the one-dimensional Weyl semimetals.

By first-principle calculations, we show that these carbon structures indeed exhibit 3D Dirac/Weyl characteristics. When cutting into slabs or nanowires in certain crystallographic directions, they can also be reduced to 2D or one-dimensional (1D) Dirac materials. A common and perhaps most important feature of the band structures for these networks is that their Weyl surfaces, lines, or points always appear near the Fermi level, making the materials truly Weyl semimetal. Moreover, the Dirac/Weyl electron characteristics are robust against external strains. A tight-binding model is developed to explain the mechanism underlying these unique electronic properties in terms of their orbital interactions. The robustness of their Weyl-type Fermi surfaces is derived from a nontrivial topology, ensured by the sublattice symmetry, which leads to an inverted band structure near the Fermi level. In addition, reasonably flat bands appear at the surfaces of these graphene networks.

The significance of these findings lies in that the 3D graphene networks provide a versatile platform for investigating the fascinating properties of Weyl electronic states. In particular, these structures may be viewed as a crystal with quasi-1D conducting channels. As such, interesting correlated electronic phases such as charge density waves, spin density waves, etc.,[30-32] may be realized, provided that the many-body correlation length can be comparable or exceed the channel separation. The interplay between single-particle Weyl electrons and these correlated electronic phases is another subject of interests. Besides the possibility of large surface ferromagnetism, the electron interaction in such systems may also lead to superconducting instabilities, and due to the huge surface density of states, this could produce a surface superconductivity with high $T_c$.[33] Last but not least, the 3D Weyl-surface semimetals may also find applications in energy storage, molecular sieves, and catalysts because of their good stability, porous geometry, and large superficial area[34-37].

## Results and Discussions

Figures 2(a-c) show an example for the three types of graphene networks, consisting of green- and purple-colored nanoribbons in Fig. 1. Looking from the top, these structures may be classified as triangular graphene network (TGN), quadrilateral graphene network (QGN), and hexagonal graphene network (HGN), respectively. By changing the width of the green nanoribbon(m) and purple nanoribbon(n), one arrives at three families of graphene networks, TGN(m,n), QGN(m,n), and HGN(m,n). The structures shown in Figs. 2(a-c) are in fact for m = n = 2. Their corresponding unit cells are shown in Figs. 2(d-f).

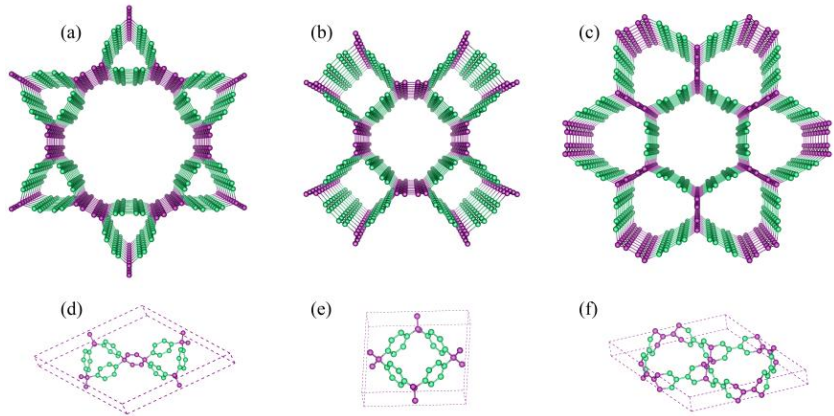

Figure 2. 3D graphene networks. (a-c) Top views of TGN(2,2), QGN(2,2), and HGN(2,2), with m = 2 and n = 2, as defined in the main text and in the caption for Fig. 1. (d-f) A tilted view of the unit cells corresponding to the graphene networks in (a-c).

Table 1 shows the structural properties for the TGN, QGN, and HGN structures, where each type has three representative structures. The corresponding properties of graphite and diamond are also shown for comparison. One can see that the bond lengths in these structures are in the range of 1.36-1.54 Å. Most of them are between those of diamond (1.54 Å) and graphite (1.42

Å). The shorter-than-1.42 Å bonds are the sp²-hybridized bonds inside the purple nanoribbons, whereas the longer-than-1.54 Å bonds are the sp³-hybridized bonds between the green and purple nanoribbons. The porous structures here lead to small carbon densities even lower than that of graphite. On the other hand, all the 3D networks have much higher bulk moduli than that of graphite. The calculated cohesive energy $E_{coh}$ indicates that these structures are all highly-stable carbon allotropes, as $E_{coh}$'s are only 0.1-0.3 eV/C smaller than those of diamond and graphite. These values in Table 1 can be contrasted to other proposed low-density carbon allotropes, such as T-carbon[38] (-6.59 eV/C), Bct-C4[39] (-7.67 eV/C), CKL[19] (-7.49 eV/C), Y-carbon[40] (-6.75 eV/C), and TY-carbon[40] (-6.71 eV/C). To be more certain on the stability, we also calculated the phonon spectra and independent elastic constants. They are shown in Fig. S1 and Table S1 of the Supplemental Material. We could not find any soft phonon mode over the entire Brillouin Zone (BZ), and all the independent elastic constants satisfy the Born stability criteria[41].

| SYSTEM | m | n | Space group | Density | Bond lengths | Bulk Moduli | $E_{coh}$ |
|---|---|---|---|---|---|---|---|
| TGN | 2 | 1 | P6₃/mmc | 1.67 | 1.42,1.45,1.52 | 201.28 | -7.70 |
|  | 2 | 2 | P6/mmm | 1.21 | 1.34-1.54 | 144.11 | -7.65 |
|  | 2 | 3 | P6₃/mmc | 0.94 | 1.36-1.54 | 112.69 | -7.66 |
| QGN | 1 | 2 | P4₂/mmc | 1.87 | 1.34-1.54 | 194.79 | -7.60 |
|  | 2 | 2 | P4/mmm | 1.51 | 1.34-1.53 | 151.46 | -7.68 |
|  | 2 | 4 | P4/mmm | 1.06 | 1.36-1.54 | 127.09 | -7.69 |
| HGN | 1 | 3 | P6₃/mcm | 1.75 | 1.36-1.54 | 208.06 | -7.61 |
|  | 2 | 2 | P6/mmm | 1.54 | 1.34-1.53 | 184.26 | -7.67 |
|  | 2 | 4 | P6/mmm | 1.24 | 1.36-1.54 | 148.05 | -7.68 |
| Diamond |  |  | Fd3̄m | 3.55 | 1.54 | 431.32 | -7.77 |
| Graphite |  |  | P6₃/mmc | 2.24 | 1.42 | 36.4 | -7.90 |

**Table 1. Structural properties for nine representative graphene networks, diamond, and graphite. They include the space group, density (g/cm3), bond lengths (Å), bulk moduli (GPa), and cohesive energy $E_{coh}$ (eV/atom).**

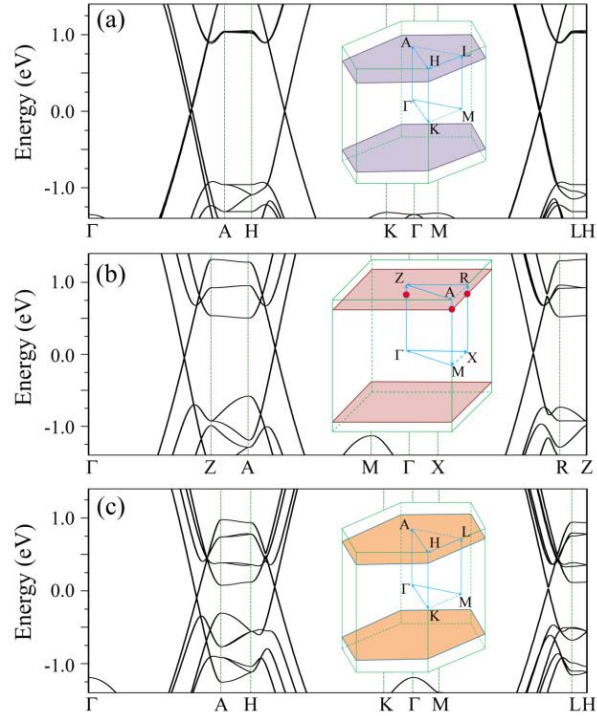

Figure 3. Band structure of graphene networks near the Fermi level (which has been set to zero): (a) TGN (2, 2), (b) QGN(2, 2), and (c) HGN(2, 2). Insets are the corresponding BZs, with the Dirac surfaces color marked. In (b), the three red dots in the inset are the positions of the Dirac points along the paths, Γ-Z, A-M, and X-R, respectively, in the band structure plot.

Figure 3 shows the band structures for TGN(2,2), QGN(2,2), and HGN(2,2), respectively. It is interesting to note that all the networks are semimetals with a Dirac linear dispersion near the Fermi level. Below, we take the simplest network, QGN(2,2) in Fig. 3(b), as an example to explain the electronic properties for all the three types. One observes from this figure that linear crossing points appear along each of the k paths, Γ-Z, A-M, and X-R. These Weyl-like points are marked on the BZ (inset of Fig. 3(b)) as three red dots, and, as a matter of fact, they are all on the reciprocal-space surface $k_z$ = 0.39 π/c, which form a Weyl surface in the BZ for QGN(2,2). By time reversal symmetry, there should be another Weyl surface at $k_z$ = -0.39 π/c. Similar Weyl surfaces are also found for TGN(2,2) and HGN(2,2). Hence, all the graphene networks are semimetals with Weyl surfaces passing their respective Fermi levels. Hence, all the 3D carbon networks studied here can be termed as Weyl-surface semimetals. It should be noted that the Weyl surfaces identified here are distinct from the conventional Fermi surface of ordinary metals: a Weyl surface is formed by a linear crossing of two electronic bands; hence the low-energy quasiparticles must be described by two-component Weyl spinors. Due to this added degeneracy, Weyl surfaces are typically not as stable as the conventional Fermi surfaces[42], (which explains why they are so rare) unless additional crystalline symmetries exist, as will be discussed in a while.

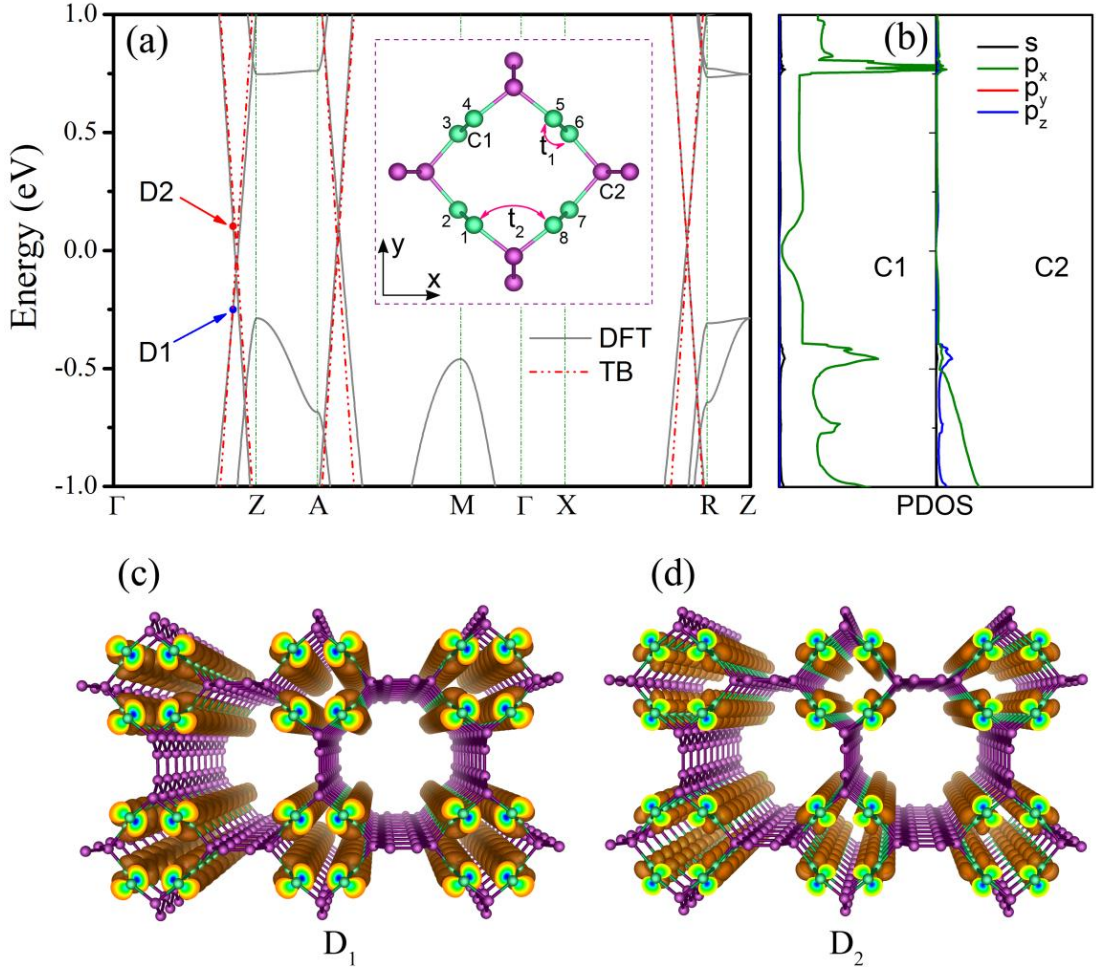

Figure 4. (a) Band structure of QGN(1,2), given by DFT (solid lines) and TB model (dashed lines). Inset is the top view of the atomic structure for QGN(1,2), where (green) C1 and (purple) C2 represent two kinds of atoms mentioned in Fig. 1. $t_1$ and $t_2$ are the intra- and inter-chain hopping energy parameters ($t_1$ = 2.95 eV and $t_2$ = 1.3 eV). (b) PDOS for the C1 and C2 atoms, respectively. (c-d) Charge densities for the states at the (blue) D1 and (red) D2 points in (a), respectively.

To understand the origin of the Weyl surfaces, we plot in Figs. 4(a-b) the band structure and partial density of states (PDOS) for QGN(1,2). The unit cell of the structure is shown in the inset of Fig. 4(a), which consists of eight green atoms labeled C1 and eight purple atoms labeled C2. As mentioned above, in the graphene network the electronic states around the Fermi level are contributed by the green-colored nanoribbons, while the function of the purple-colored nanoribbons is to provide the backbones of the structures. Therefore, the PDOS in Fig. 4(b) reveals that the Weyl bands around the Fermi level are originated from states only on C1 atoms, with the relevant electron orbitals being $p_x$ or $p_y$, rather than being $p_z$ or $s$. As a comparison, Figs. 4(c-d) show the charge density profiles for the states in the Weyl bands. One can see that the charge density on a C1 atom corresponds to that of one π-orbital, which is very similar to that in an armchair carbon nanotube. Because the purple-colored zigzag nanoribbons have little contribution to the Weyl bands, we can thus omit them and, in the following discussion, just consider the green-colored zigzag nanoribbons, which in a unit cell form a closed cylinder just like a distorted armchair carbon nanotube. In this sense, the graphene network can be viewed as a 3D bundle of carbon nanotubes. It is known that the nanotubes are 1D Weyl semimetals with linearly crossing Fermi points in their respective band structures[43]. A simple physical picture is that, when forming the carbon networks, the inter-tube coupling between the σ electrons in the x-y plane is rather strong, leading to significant dispersions, but the inter-tube coupling between the π electrons in the same plane is vanishingly small, leading to the Weyl surfaces which are almost dispersionless in the x-y plane.

Because only the π orbitals of the C1 atoms contribute to the Weyl surface properties of QGN(1,2), a tight-binding (TB) model, based on a single orbital per C1 site, can be used to describe its properties around the Fermi level,

$$H = \sum_{<i,j>} \sum_\mu t_{ij} e^{-i\mathbf{k}\cdot \mathbf{d}_{ij}^\mu}, \qquad (1)$$

where $i,j \in \{1,2,3...8\}$ are the eight C1 sites as in the inset of Fig. 4(a), $\mathbf{d}_{ij}^\mu$ is a vector directed from $j$ to $i$, $t_{ij}$ is the hopping energy between $i$ and $j$, and $\mu$ runs over all equivalent lattice sites under translation. For $t_{ij}$, we only consider two hopping processes: one is the nearest-neighbor interaction inside one zigzag nanoribbon $t_1$; the other is between the nearest sites of two neighboring green nanoribbons $t_2$, as indicated in the inset of Fig. 4(a). The spectrum of energy band is symmetric about zero energy because of the presence of a sublattice symmetry with two sublattices of sites {1,3,5,7} and {2,4,6,8} respectively. Equation (1) can be easily diagonalized to yield

$$(\lambda - x_0)(\lambda + x_1)^2(\lambda - x_2) = 0, \qquad (2)$$

where $\lambda = E^2$, $x_0 = [2t_1 \cos(k_c c/2) + t_2]^2$, $x_1 = -[4t_1^2 \cos^2(k_c c/2) + t_2^2]$ and $x_2 = [2t_1 \cos(k_c c/2) - t_2]^2$. It is straight forward to find that zero-energy states would appear if $x_0 = 0$ or $x_2 = 0$, i.e., $\cos(k_c c/2) = \pm \frac{t_2}{2t_1}$. Because we should have $t_2 < t_1$, the zero-energy states appear on two separate surfaces at $k_z = K_c \equiv \pm(2/c)\arccos[\sqrt{t_2/(2t_1)}]$. Despite its simplicity, the model captures the essential physics of the first-principles results. In Fig. 4(a), the fitted band structure of the tight-binding model is also shown, which agrees rather well with first-principles results. Around the Weyl surface at $\tau_z K_c$ ($\tau_z = \pm 1$), the low-energy quasiparticles are described by the effective Hamiltonian

$$H_{\text{eff}}(q_z) = \tau_z v q_z \sigma_z, \qquad (3)$$

where $q_z = k_z - K_c$ is the wave vector component normal to the Weyl surface, $v \approx 1.0 * 10^6$ m/s is the Fermi velocity, and the Pauli matrix $\sigma_z$ denotes the two bands crossing at the surface. For each surface, Eq. (3) takes the form of a 1D Weyl Hamiltonian, and the chirality reverses between the two surfaces as required by the time reversal symmetry. These are the essential features of the Weyl-surface semimetals.

As mentioned, the Weyl surface is in general not stable unless with symmetry/topology protection. For the structures studied here, the stability is closely related to their sublattice symmetry, which is ensured, e.g., by the crystalline mirror symmetries inherent to the structure along x, y, or a combination of the two. With negligible spin-orbit coupling for carbon, such systems fall into the BDI topological class with a zero-dimensional $\mathbb{Z}_2$ topological invariant defined at any point in the BZ with a local gap[44-46]. The $\mathbb{Z}_2$ invariant just indicates whether the gap is inverted or not, with reference to the normal band ordering in the atomic limit. In the graphene networks, the band gap is inverted ($\mathbb{Z}_2 = 1$) near the central region of BZ while it is un-inverted ($\mathbb{Z}_2 = 0$) around Z-point at the BZ boundary, hence the Weyl surfaces which separates the two regions with different band topologies cannot be gapped as long as the sublattice symmetry is maintained.

We find that under both uniaxial and biaxial strains, the mirror symmetry hence the sublattice symmetry is preserved for the graphene networks. Therefore, when a uniaxial/biaxial compressive strain is applied to the structure, the Weyl surfaces simply shift their positions to the center of the BZ; when a uniaxial/biaxial tensile strain is applied, on the other hand, the Dirac surfaces will shift in the opposite direction (see Fig. S2 of the Supplemental Material). Only when the sublattice symmetry is broken, the Dirac surfaces would be destroyed so the system are gapped or become metallic. However, this symmetry breaking is difficult because the networks are strongly against such distortions, as revealed by their share moduli in Table S1. Therefore, the Weyl surfaces in the graphene networks are quite robust.

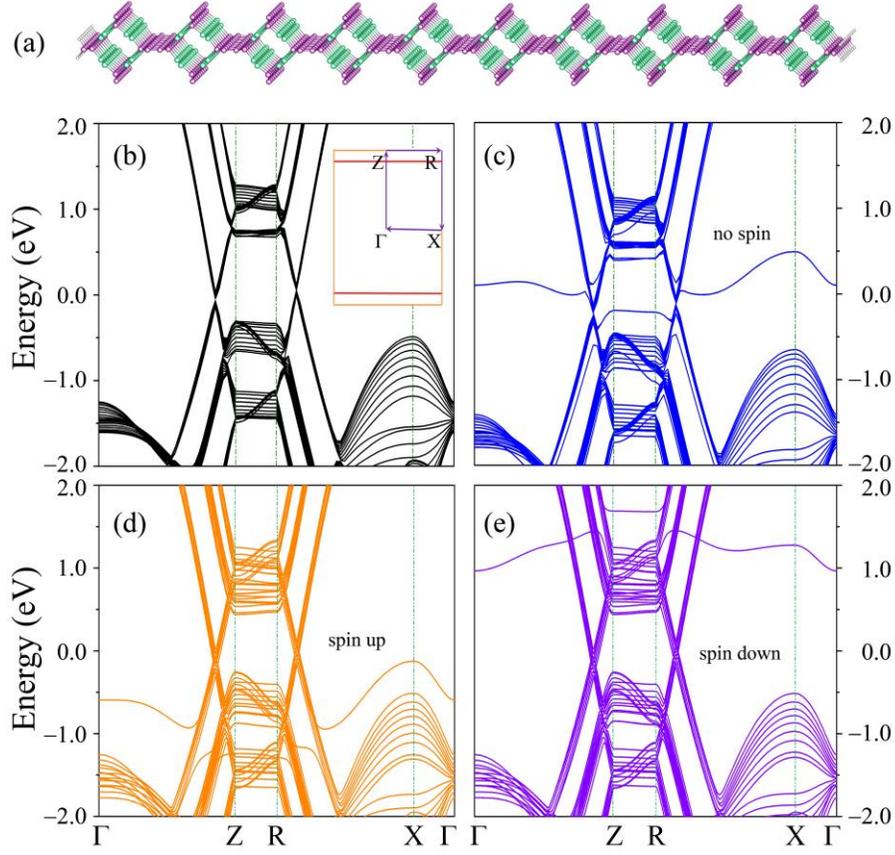

Figure 5. (a) Atomic structure of a 10-layer thick slab of QGN(1,2). Each atom on the surfaces is passivated by two hydrogen atoms. (b) Corresponding band structure. Inset shows the BZ where two red lines represent the two Dirac lines. (c) Non-spin-polarized band structure of the 10-layer thick slab when each atom on the surfaces is passivated by only one hydrogen atom. (d-e) Corresponding spin-polarized band structures for (c).

When the 3D graphene network is cut in parallel to the zigzag-chain directions, a slab will be obtained. Figure 5(a) shows a 10-layer slab for QGN(1,2), where for the ease of discussion, each carbon atom on the surfaces is passivated by two hydrogen atoms. Figure 5(b) shows the band structure of the slab, from which one see two linearly crossing points at the Fermi level. A closer examination reveals that the cut leads to two Weyl lines on the 2D BZ, as shown in the inset of Fig. 5(b). Therefore, the slab is also a Weyl-line semimetal, whose Weyl lines can be understood either as a projection of the Weyl surfaces from a 3D BZ to a 2D BZ, or as an expansion of the 1D Weyl points(of isolated nanotubes) in a 1D BZ to a 2D BZ. Here, the purpose of passivation by hydrogen is only to eliminate the dangling bonds at the surfaces. It should not have any effect on the physics of most of the Weyl electrons, which mainly reside inside the slab. It is, however, interesting what might happen when each carbon atom on the surfaces is passivated by only one hydrogen atom, which would expose one of the surface states when there is no hydrogen passivation. Figure 5(c) shows the corresponding band structure, calculated without spin polarization. A rather-flat and half-occupied surface band appears that connects the two Weyl points near the Fermi level. It is known that, in a flat band, there could be strong electron correlation effects[47-50]. Figures 5(d-e) shows that, when the electron spin is considered, the flatband splits into a spin-up and a spin-down bands, leading to surface ferromagnetism. Interestingly, the Weyl electrons do not participate in the magnetism, nor a majority of them are affected by the presence of the surface ferromagnetism due to their strong localization within individual nanotubes.

When the 2D slabs are being further cut, 1D nanowires are obtained. Similar to the 2D slabs, the electronic properties of the nanowires here also depend on the surface passivation. Figure 6(a) shows the calculated band structure when each surface atom is passivated by two hydrogen atoms. Clearly, the nanowires are 1D Weyl-point semimetals. Figure 6(b) shows the non-spin polarized band structure, when each surface atom is passivated by only one hydrogen atom. In this case, a number of rather flat surface and near-surface bands appear near the Fermi level. As expected, Figs. 6(c-d) reveals that the spin-polarization has a significant effect on these bands, again leading to surface ferromagnetism.

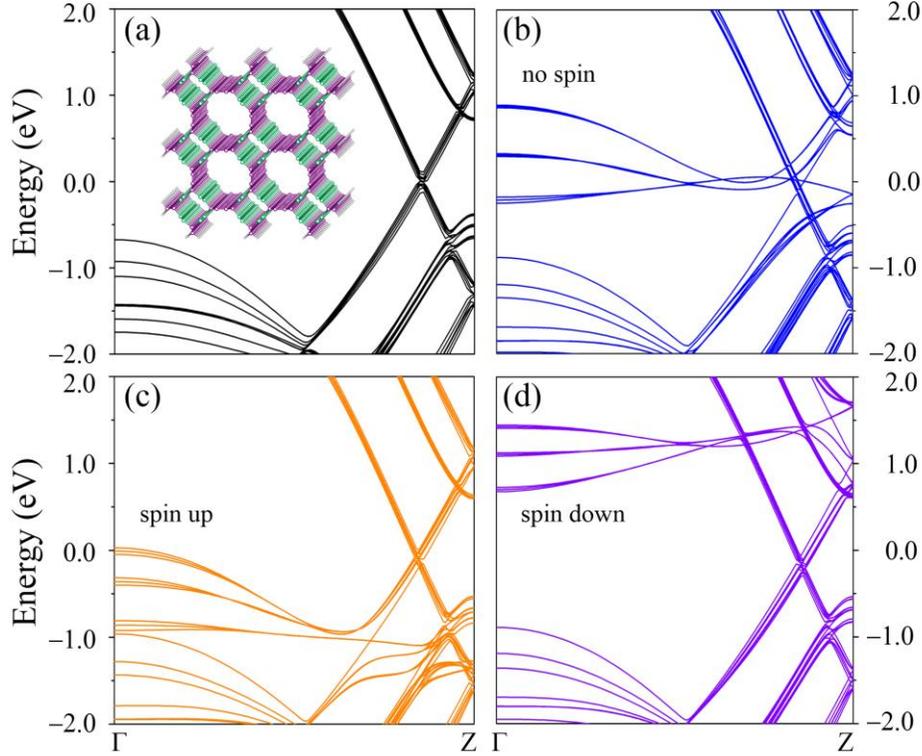

Figure 6. Band structure for the nanowire obtained by cutting a 3×3 unit cell out of QGN(1,2). (a) When each atom on the surfaces is passivated by two hydrogen atoms. Inset shows the atomic structure. (b) When each atom on the surfaces is passivated by only one hydrogen atom (non-spin-polarized). (c-d) Same as in (b) but with spin polarization.

The Weyl lines/points in the 2D/1D graphene networks as well as the appearance of flat surface bands are also derived from the topological properties of the bulk band structure. For example, similar to the Weyl surfaces, the surfaces of the 3D network structures also represent an interface separating the gap-inverted nontrivial region and the (trivial) vacuum; hence there exist surface bands in between the projected Weyl surfaces on the side surfaces, as observed in Fig. 5(c). And because the sublattice symmetry may not be respected at the system boundary, the surface bands could be distorted from being flat or even pushed away from the Fermi level depending on the boundary conditions, in analogy with the edge states of graphene[51].

## Conclusions

We have examined three families of 3D graphene networks by first-principles calculations. All of them turn out to be 3D Weyl-surface semimetals with two almost flat Weyl surfaces at the Fermi level in their respective band structures. The characteristics of these Weyl surfaces may be traced back to the topology and symmetry of the underlying structures. In particular, the graphene networks are a 3D bundle of zero-gap carbon nanotubes, with inherent 1D Weyl electron conduction channels, connected by gapped carbon nanoribbons. One may cut the graphene networks to obtain 2D slabs or 1D nanowires. They are also Weyl-type semimetals. If each carbon atom on the surfaces is only passivated by one hydrogen atom, nearly flat surface bands appear and the surfaces become ferromagnetic due to strong Coulomb repulsion between localized electrons. A tight-binding model was constructed, which allows us to explain most of the silent features predicted by first-principles calculations.

# Methods

We performed first-principles calculations within the density functional theory (DFT) as implemented in the VASP codes[52]. The potential of the core electrons and the exchange-correlation interaction between the valence electrons were described, respectively, by the projector augmented wave[53] and the generalized gradient approximation (GGA) with Perdew-Burke-Ernzerhof (PBE) functional[54]. A kinetic energy cutoff of 500 eV was used. The atomic positions were optimized using the conjugate gradient method, and the energy and force convergence criteria were set to be $10^{-5}$ eV and $10^{-2}$ eV/Å, respectively. In the calculations of the slab geometry, the two nearest slabs were separated by a vacuum layer of at least 10 Å to avoid artificial interactions. To sample the BZ, we used the k-point grids with a spacing = $2\pi \times 0.02$ Å$^{-1}$ for tetragonal systems and the Γ-centered sampling scheme for hexagonal systems, respectively, within the Monkhorst-Pack sampling scheme[55].

**Corresponding Author**

*E-mail address:
chenyp@xtu.edu.cn (Y.Chen);
xieyech@xtu.edu.cn (Y. Xie);
zhangs9@rpi.edu (S. Zhang).

**ACKNOWLEDGMENT**

This work was supported by the National Natural Science Foundation of China (Nos. 51176161, 51376005 and, 11474243) and the Hunan Provincial Innovation Foundation for Post-graduate (No. CX2015B211). SAY was supported by SUTD-SRG-EPD2013062. MLC was supported by the Theory Program at the Lawrence Berkeley National Lab through the Office of Basic Energy Sciences, U.S. Department of Energy under Contract No. DE-AC02-05CH11231, and by the National Science Foundation under Grant No.DMR15-1508412.SZ acknowledges support by US DOE under Grant No.DE-SC0002623.